\documentclass[preprint,12pt]{elsarticle}
\usepackage{amssymb}
\usepackage{amsmath}
\usepackage{graphicx}
\usepackage{float}

\usepackage{caption}
\usepackage[colorlinks=true, allcolors=blue]{hyperref}
\makeatletter
\def\ps@pprintTitle{%
  \let\@oddhead\@empty
  \let\@evenhead\@empty
  \let\@evenfoot\@empty
  \def\@oddfoot{\hfill\@date\hfil} 
  \let\@mkboth\@gobbletwo
  \def\thepage{\arabic{page}}
}
\makeatother

\begin{document}

\begin{frontmatter}

\title{Physics-Informed Neural Networks for High-Precision Grad-Shafranov Equilibrium Reconstruction}

\author[1]{Cuizhi Zhou}
\author[1]{Kaien Zhu\corref{cor1}}

\ead{zhukaien@nankai.edu.cn}
\cortext[cor1]{Corresponding author}

\affiliation[1]{
  organization={College of Physics, Nankai University},
  addressline={94 Weijin Rd},
  city={Tianjin},
  postcode={300071},
  country={China}
}

\begin{abstract}

The equilibrium reconstruction of plasma is a core step in real-time diagnostic tasks in fusion research. This paper explores a multi-stage Physics-Informed Neural Networks(PINNs) approach to solve the Grad-Shafranov equation, achieving high-precision solutions with an error magnitude of $O(10^{-8})$ between the output of the second-stage neural network and the analytical solution. Our results demonstrate that the multi-stage PINNs provides a reliable tool for plasma equilibrium reconstruction.

\end{abstract}


\end{frontmatter}

\section{Introduction}

With the ever-increasing global energy demand, traditional energy sources struggle to sustain humanity's long-term needs. Renewable sources like wind and solar, while promising, are too intermittent to fully replace conventional energy. Nuclear fusion, however, offers a promising solution due to its high energy yield and clean production process, positioning it as a potential future mainstream energy source \cite{Smith2010}. Plasma equilibrium is a critical step in achieving controlled nuclear fusion. Therefore, efficiently and accurately solving the Grad-Shafranov (GS) equation \cite{Grad1958,Shafranov1958,Lao2005} can accelerate the realization of controlled fusion and help address the global energy challenge.

The GS equation is a high-order nonlinear partial differential equation that often lacks analytical solutions. Traditional numerical methods for solving the GS equation, such as finite element methods \cite{ALBANESE2023108804,RICKETSON2016744,HEUMANN2017522}, spectral element methods \cite{LI2021107264,PALHA201663,HOWELL20141415}, and hybrid finite  volume /finite element methods \cite{FAMBRI2023112493,ZAMPA2024346}, have achieved significant progress. However, these approaches still face challenges in handling noisy data, generating complex meshes, and solving inverse problems efficiently \cite{Karniadakis2021}. PINNs \cite{raissi2017physics,RAISSI2019686,Karniadakis2021} offer distinct advantages, including mesh-free computation, flexibility in handling complex boundary and initial conditions, simultaneous solution of forward and inverse problems, and seamless integration of data with physical models. These strengths have led to extensive applications of PINNs in solving partial differential equations (PDEs).

In 1995, researchers first explored using neural networks to solve the GS equation \cite{PhysRevLett.75.3594}. The core idea was conceptually similar to today's PINNs, but due to technological limitations at the time, derivatives were computed using manually derived methods. In recent years, with the rapid advancement of neural networks, the fusion community has increasingly employed neural networks as surrogate models \cite{joung2019deep,merlo2021proof,wai2022neural,Liu_2022} for solving the GS equation. However, these approaches have not yet fully leveraged the PINN framework.

With the rapid advancement of PINNs, they have been increasingly applied to solve the GS equation. H. Baty \cite{10.1093/mnras/stad3320} employed the basic PINN framework to solve the GS equation, achieving a maximum error of $O(10^{-3})$ compared to analytical solutions. B. Jang \cite{jang2024grad} compared the performance of hard-constrained and soft-constrained PINNs in solving the parameterized GS equation (with adjustable parameters including the inverse aspect ratio $\epsilon $, triangularity $\lambda $, and elongation $b$ ). Both methods demonstrated errors on the order of $O(10^{-4})$ when compared to analytical solutions. F. N. Rizqan \cite{rizqan2025evaluation} combined Fourier Neural Operators (FNOs) with PINNs to solve the parametric GS equation, reporting average errors of $O(10^{-2})$ in both training and extrapolation test regions.

Due to the inherent limitations of neural networks, it is challenging for PINNs to achieve extremely high precision in solving the GS equation. Currently, few existing PINN-based methods for solving the GS equation have been able to reduce the error compared to analytical solutions to below the order of $O(10^{-5})$ .

In this study, we adopt the multi-stage PINNs framework proposed by Y. Wang in 2024 \cite{WANG2024112865} to solve the GS equation, achieving high-precision solutions. We use two-stage PINNs. The trained first-stage neural network reduces the error compared to the analytical solution to the order of $O(10^{-4})$ , while the second-stage neural network further improves the accuracy, achieving an error of $O(10^{-8})$ .

\section{High-precision solution of the Grad-Shafranov equation via Physics-Informed Neural Networks}

The equilibrium state of plasma is a critical step for achieving net energy output in fusion reactions. Therefore, the accurate and rapid solution of the GS equation holds great significance for research in the field of nuclear fusion \cite{WU2024}.

Derived from the ideal magnetohydrodynamic (MHD) equilibrium equations, the GS equation, that is, 
\begin{equation}
    \Delta^{\star} \psi = \frac{\partial^2 \psi}{\partial Z^2} + R \frac{\partial}{\partial R}(\frac{1}{R} \frac{\partial\psi}{\partial R}) = -\mu_0 R^2 \frac{dP}{d\psi} - g\frac{dg}{d\psi},
    \label{gs_equation}
\end{equation}
is the core equation describing the equilibrium of magnetically confined fusion plasmas. It is used to calculate the poloidal flux function $\psi(R,Z)$ of the plasma, thereby determining the magnetic surface configuration, current distribution, and pressure distribution.  We use  cylindrical coordinates $(R,Z,\phi)$ and $\mu_0$ is the vacuum permeability, $P(\psi)$ is the plasma pressure distribution, $g(\psi) = RB_\phi$ is the toroidal magnetic field function. The boundary is named as
\begin{equation}
    \psi = \mathcal{G}(R,Z),  (R,Z) \in \partial \Omega.
\end{equation}
When the boundary is the last closed flux surface (LCFS), $ \mathcal{G}(R,Z) = 0$ .

\subsection{Multi-stage Physics-Informed Neural Networks}

Neural networks are prone to failure modes that often trap the training process in local minima\cite{NEURIPS2021_df438e52}, leading to a stagnant equilibrium after a certain number of iterations. Existing solutions rarely achieve errors below $O(10^{-5})$ . To address this, Yongji Wang proposed a multi-stage neural network approach in 2024 \cite{WANG2024112865} that can nearly reach double-precision machine accuracy of $O(10^{-16})$. The core idea is to train successive neural networks, each targeting the residual errors from the previous stage until satisfactory precision is achieved.

\begin{figure}
    \centering
    \includegraphics[width=0.9\linewidth]{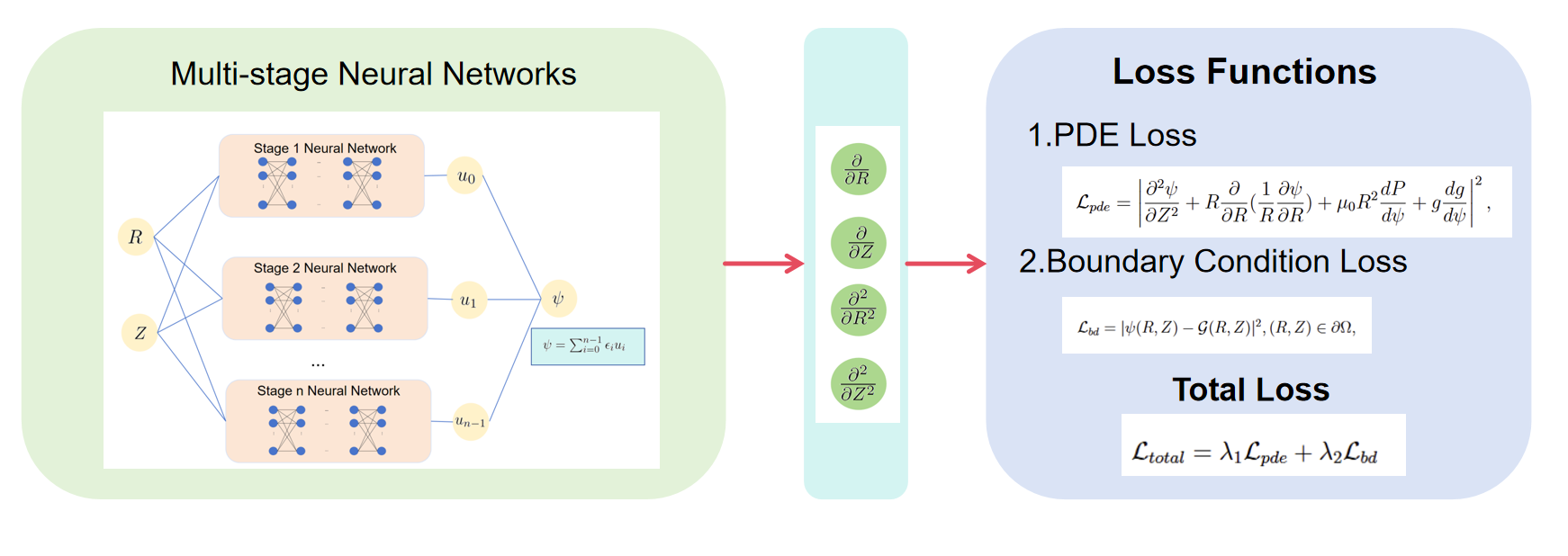}
    \caption{Multi-stage PINNs architecture for Grad-Shafranov equation.}
    \label{fig:PINN structure}
\end{figure}

Figure \ref{fig:PINN structure} illustrates the neural network architecture of the multi-stage PINNs for solving the GS equation. When applying the multi-stage PINNs to solve the GS equation, neural networks are trained sequentially stage by stage. The loss functions are
\begin{equation}
    \mathcal{L}_{pde} = \left| \frac{\partial^2 \psi}{\partial Z^2} + R \frac{\partial}{\partial R}(\frac{1}{R} \frac{\partial\psi}{\partial R}) + \mu_0 R^2 \frac{dP}{d\psi} +  g\frac{dg}{d\psi}\right|^2,
\end{equation}
\begin{equation}
    \mathcal{L}_{bd} = |\psi(R,Z) - \mathcal{G}(R,Z)|^2,   (R,Z) \in \partial \Omega,
\end{equation}
\begin{equation}
    \mathcal{L}_{total} = \lambda_1 \mathcal{L}_{pde} + \lambda_2\mathcal{L}_{bd},
\end{equation}
and 
\begin{equation}
    \psi = \sum_{i = 0} ^{n-1} \epsilon_iu_i.
\end{equation}
The first-stage neural network $u_0$ is trained until the total loss $\mathcal{L}_{total}$ stabilizes, after which the second-stage neural network is trained, and this process is repeated for subsequent stages. During training, the coefficients $\lambda_1$  and $\lambda_2$ in the loss function of each stage's PINN can be adjusted differently according to specific requirements. $\epsilon$ in each stage's neural network represents a normalization coefficient: for the first-stage neural network, the normalization coefficient $\epsilon_1$ is generally set to 1, while the normalization coefficients for subsequent multi-stage neural networks are determined based on the residual errors from the previous stage. For detailed theoretical derivations, refer to Yongji Wang's 2024 literature \cite{WANG2024112865}.

\subsection{Two-stage PINNs solution for the GS Equation}

In most cases, the GS equation does not have an analytical solution, and the functions $P(\psi)$ and $g(\psi)$ in the equation are obtained by solving the base-state equation using numerical methods. To demonstrate the high-precision solving capability of PINNs, we selected three typical cases with analytical solutions from the literature to investigate the errors of PINNs. In this paper, the two-stage Physics-Informed Neural Networks (PINNs) have been well-trained, achieving errors on the test set at the order of $10^{-7}$ and $10^{-8}$ compared with the analytical solutions. Given that such accuracy is satisfactory, more advanced-stage PINNs have not been employed.

\subsubsection{PINNs solution for Solov\'{e}v equilibrium}

\begin{figure}
    \centering
    \includegraphics[width=0.5\linewidth]{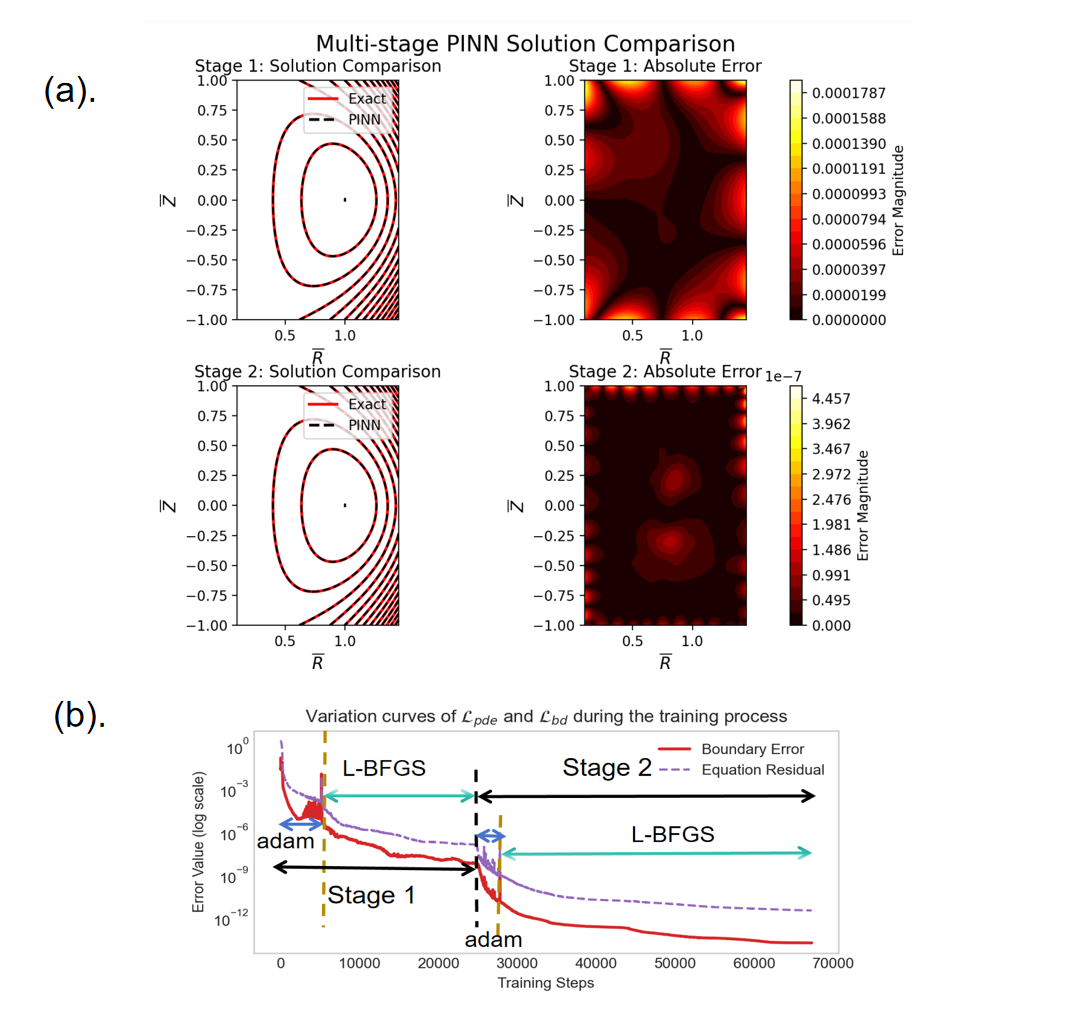}
    \caption{Solution results of the Solov\'{e}v equilibrium via Physics-Informed Neural Networks. (a) Solution results of multi-stage neural PINNs (left) and errors compared with analytical solutions (right). (b) Variation curves of $\mathcal{L}_{pde}$ and $\mathcal{L}_{bd}$ during the training process.}
    \label{fig:rectangular_device}
\end{figure}

The first example is Solov\'{e}v equilibrium,
\begin{equation}
    \frac{dP}{d\psi} = - \frac{c_1}{\mu_0},
\end{equation}
\begin{equation}
    g\frac{dg}{d\psi} = -c_2R_0^2.
\end{equation}
Here, $c_1$, $c_2$ and $R_0$ are constants, and the analytical solution in this case is:
\begin{equation}
    \psi = \frac{1}{2}(c_2R_0^2+c_0R^2)Z^2+\frac{1}{8}(c_1-c_0)(R^2-R_0^2)^2.
\end{equation}
In this paper, the parameters from Youjun Hu's Notes on Tokamak Equilibrium \cite{Hu_NotesTokamak} are adopted as the constants of the equation, specifically $c_0 = B_0/(R_0^2\kappa_0q_0)$,$c_1 = B_0(\kappa_0^2+1)/(R_0^2\kappa_0q_0)$,$c_2 = 0$. The analytical solution of the equation under these parameters is:
\begin{equation}
    \psi = \frac{B_0}{2R_0^2\kappa_0q_0}[R^2Z^2+\frac{\kappa_0^2}{4}(R^2-R_0^2)^2].
    \label{eq:exact_solution}
\end{equation}
Define $\psi_0 = B_0R_0^2$ and $\overline{\psi} = \psi/\psi_0$. Equation (\ref{eq:exact_solution}) then becomes:
\begin{equation}
    \overline{\psi} = \frac{1}{2\kappa_0q_0}[\overline{R}^2\overline{Z}^2  + \frac{\kappa_0^2}{4}(\overline{R}^2-1)^2].
\end{equation}
Here, $\overline{R} = R/R_0$ and $\overline{Z} = Z/R_0$. The corresponding normalized GS equation is:
\begin{equation}
    \frac{\partial^2\overline{\psi}}{\partial \overline{Z}^2} + \overline{R}\frac{\partial}{\partial \overline{R}}(\frac{1}{\overline{R}}\frac{\partial \overline{\psi}}{\partial \overline{R}}) = \overline{R}^2\frac{\kappa_0^2+1}{\kappa_0q_0},
    \label{eq:PINN_eq1}
\end{equation}
\begin{equation}
    \overline{\psi} = \mathcal{G}(R,Z) ,(R,Z) \in \partial \Omega
    \label{eq:rectangular_bd}.
\end{equation}
Here, the constants are set as $\kappa_0 = 1.5$ and $q_0 = 1.5$. Equation (\ref{eq:PINN_eq1}) is the PDE to be solved, with the solution domain being $R\in[0.1,1.45] \times Z \in [-1,1]$. Equation (\ref{eq:rectangular_bd}) represents the boundary conditions, where the points on the boundary in this paper are given by the analytical solution.

In this paper, a two-stage PINNs is employed to solve the GS equation over a rectangular domain. In this example, each stage of the neural network consists of three hidden layers, with 20 neurons in each hidden layer, resulting in a total of 921 trainable parameters. The hyperbolic tangent function is used as the activation function in the neural network, with a learning rate of $10^{-3}$ and a batch size set to 100. The method for generating collocation points within the solution domain is Latin hypercube random sampling \cite{loh1996latin}. For the training of the first-stage neural network, the number of collocation points is 2001. In each stage, the Adam optimizer \cite{Kingma2014AdamAM} and the L-BFGS optimizer \cite{LiuNocedal1989} are used for training in sequence. When switching optimizers, the collocation points are regenerated to avoid overfitting. The number of collocation points for the training of the second-stage neural network is 4002.

After the final training of the neural network, the comparison between the results of the first-stage neural network in solving the GS equation and the analytical solution shows that the error can reach the order of $O(10^{-4})$, while the error of the two-stage neural network can reach the order of $O(10^{-7})$, as shown in Fig. \ref{fig:rectangular_device}(a). In this example, the test set is different from the training set. The test set consists of 100×100 uniformly distributed grid points within a rectangular region. Figure \ref{fig:rectangular_device}(b) presents the solution results of the multi-stage PINNs and the variation curves of $\mathcal{L}_{\text{pde}}$ and $\mathcal{L}_{\text{bd}}$ during the training process. It can be clearly observed that the use of the second-stage neural network allows the loss function to continue decreasing after the training of the first-stage neural network has stabilized.After the final training, $\mathcal{L}_{pde}$ is reduced to the order of $10^{-11}$, and $\mathcal{L}_{bd}$ is reduced to the order of $10^{-13}$. 

\subsubsection{Solving the GS equation in a droplet-shaped domain}

\begin{figure}
    \centering
    \includegraphics[width=0.5\linewidth]{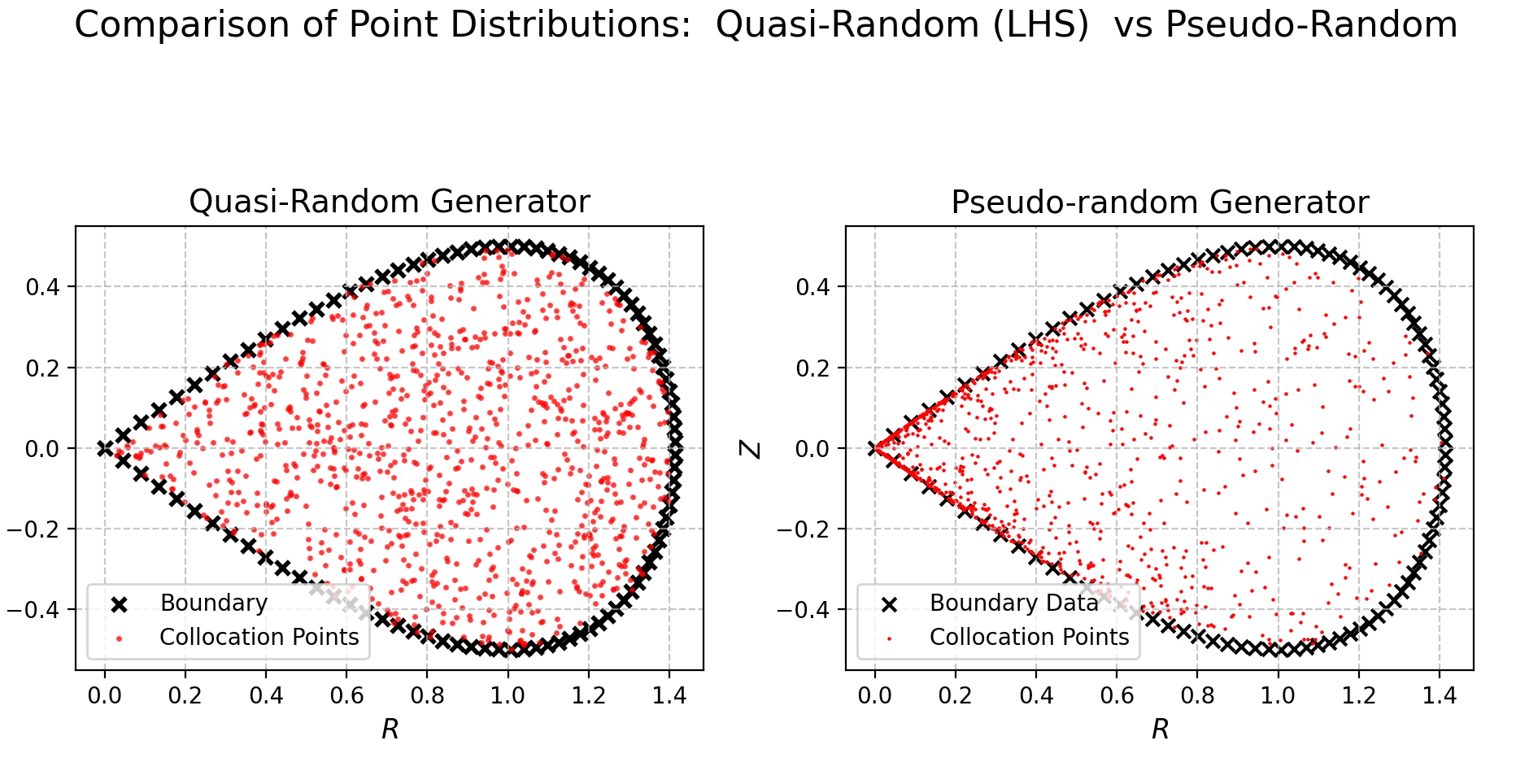}
    \caption{Comparison of point distributions:  Latin hypercube method vs pseudo-random generator.}
    \label{fig:drop-collcation}
\end{figure}

Different forms of the right-hand side of the GS equation result in solutions with different shapes \cite{deriaz2011magnetic}. When the form of the GS equation is as shown in Eq. (\ref{eq:drop_eq}),
\begin{equation}
     \frac{\partial^2 \psi}{\partial Z^2} + R \frac{\partial}{\partial R}(\frac{1}{R} \frac{\partial\psi}{\partial R}) = -f_0(R^2+R_0^2),
     \label{eq:drop_eq}
\end{equation}
the corresponding analytical solution is:
\begin{equation}
    \psi(R,Z) = \frac{f_0 R_0 ^2}{2}(a^2 -Z^2 - \frac{(R^2-R_0^2)^2}{4R_0^2}).
\end{equation}
On the boundary $\partial\Omega$, $\psi = 0$,
\begin{equation}
    \partial \Omega = \{R = R_0 \sqrt{1+\frac{2acos(\alpha)}{R_0}},z = aR_0,\alpha = [0: 2\pi]\}.
\end{equation}
Here, the constants are set as $f_0 = 1$, $a = 0.5$, and $R_0 = 1$.

In this paper, collocation points within the rectangular domain are generated through random sampling using the Latin hypercube method \cite{loh1996latin}, while collocation points in the boundary region are generated by means of a masking approach. Different from the collocation point distribution generated by the pseudo-random generator used by Hubert Baty \cite{2024arXiv240300599B}, the Latin hypercube method has stronger randomness. Figure \ref{fig:drop-collcation} shows the effect of collocation points generated by the two different generators, with 1000 boundary points and 100 internal collocation points. The neural network structure and parameters used in this example are the same as those in the previous example. The test set consists of 1000×1000 uniformly distributed grid points within a rectangular region, with points outside the boundary area removed by means of masking.
\begin{figure}
    \centering
    \includegraphics[width=0.5\linewidth]{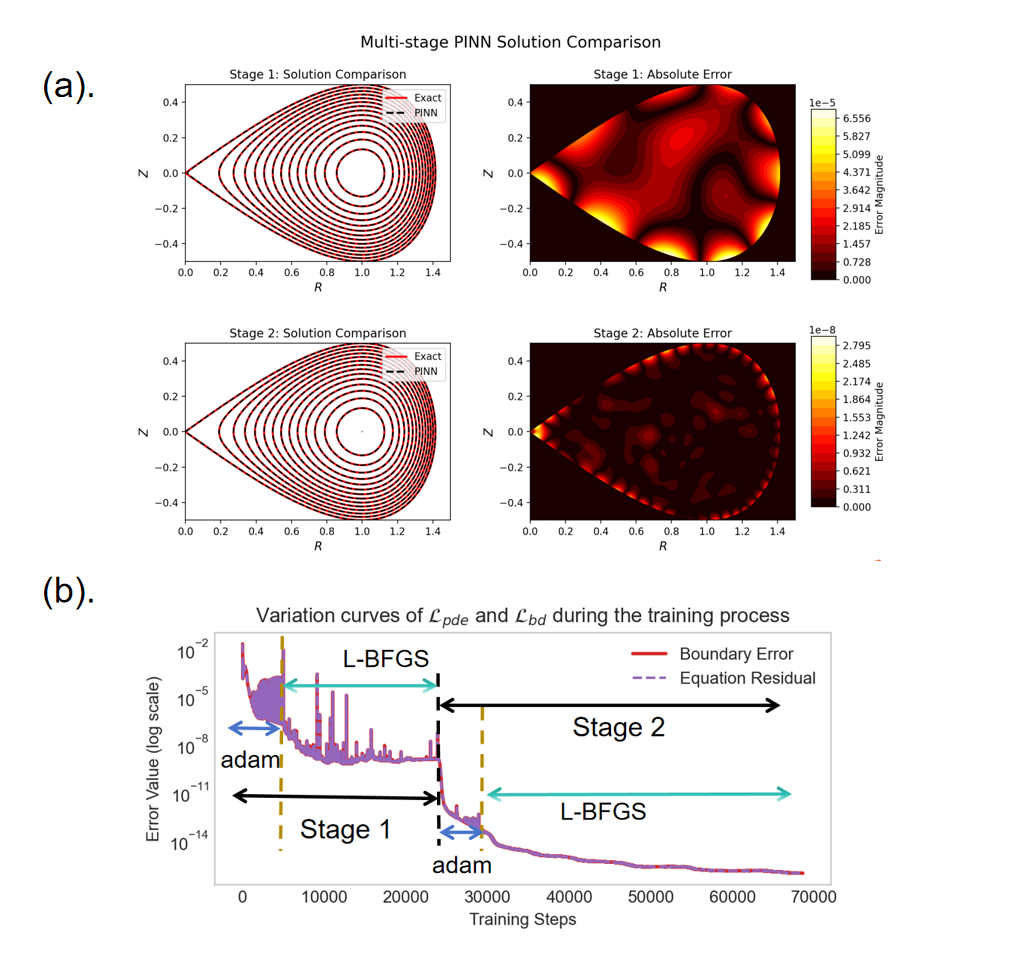}
    \caption{Solution results of the Grad-Shafranov equation in drop domain. (a) Solution results of multi-stage neural PINNs (left) and errors compared with analytical solutions (right). (b) Variation curves of $\mathcal{L}_{pde}$ and $\mathcal{L}_{bd}$ during the training process.
}
    \label{fig:drop_device}
\end{figure}

Figure \ref{fig:drop_device}(a) shows the results of solving the GS equation using PINNs for the droplet-shaped device, as well as the corresponding error compared with the analytical solution. After training with the first-stage neural network, the error relative to the analytical solution for this device can reach the order of $O(10^{-5})$. On the basis of the first-stage neural network, after training with the second-stage neural network, the error can reach the order of $O(10^{-8})$, which is a reduction of three orders of magnitude. Figure \ref{fig:drop_device}(b) presents the variation curves of $\mathcal{L}_{\text{pde}}$ and $\mathcal{L}_{\text{bd}}$ throughout the training stage. After the training is completed, both $\mathcal{L}_{pde}$ and $\mathcal{L}_{bd}$ are reduced to the order of $O(10^{-15})$. This paper uses this example to demonstrate the capability of the multi-stage PINNs in solving the GS equation in the presence of an X-point. It can be seen that the X-point can be clearly identified when the second-stage neural network is used.

\subsubsection{Solov\'{e}v  equilibrium with triangularity parameter}

When describing the equilibrium solutions of tokamaks, triangular parameters are sometimes used. In this paper, the example mentioned in \cite{deriaz2011magnetic} is cited to verify that the multi-stage neural network can achieve high-precision solving of the GS equation under such circumstances.

Define $R = R_0 + a x = R_0(1+\epsilon x)$ and $Z= ay$. The expression of the GS equation under this definition is:
\begin{equation}
     -\Delta^{\star} \psi = \alpha (R_0(1+\epsilon x))^2 + \beta,
     \label{eq:toroidal_eq}
\end{equation}
where $\alpha = \frac{4(a^2 + b^2)\epsilon + a^2(2 \lambda - \epsilon ^3)}{2R_0 ^2 \epsilon a^2 b^2}$, $\beta = - \frac{\lambda}{b^2\epsilon}$, and $ \Delta ^{\star} = \frac{\partial ^2 \psi}{\partial x ^2}+ \frac{\partial^2\psi}{\partial y ^2} - \frac{\epsilon}{1+ \epsilon}\frac{\partial \psi}{\partial x}$.
When the GS equation takes the form shown in Eq. (\ref{eq:toroidal_eq}), the corresponding analytical solution is given by:
\begin{equation}
    \psi(x,y) = 1-(x-\frac{\epsilon}{2}(1-x^2))^2 - ((1-\frac{\epsilon ^2}{4})(1+\epsilon x)^2 + \lambda x(1+ \frac{\epsilon}{2}x))(\frac{a}{b}y)^2
    \label{eq:toroidal_exact}
\end{equation}
Considering the boundary defined as:
\begin{equation}
    \partial \Omega = \{(x, y) \in \mathbb{R}^2 \mid \psi(x, y) = 0\},
\end{equation}
the explicit expression of the boundary condition is:
\begin{equation}
    y = \pm \frac{b}{a} \sqrt{\frac{1 - \left(x - \frac{\varepsilon}{2}(1 - x^2)\right)^2}{\left(1 - \frac{\varepsilon^2}{4}\right)(1 + \varepsilon x)^2 + \lambda x \left(1 + \frac{\varepsilon}{2}x\right)}}, \quad x \in [-1, 1].
\end{equation}

We employ the PINNs to solve the GS equation under these conditions. In this scenario, the neural network takes the $(x, y)$ coordinates as input and outputs the magnetic flux $\psi$.The parameters in the equation are set as follows: the inverse aspect ratio $\epsilon = 3/10$, the triangularity $\lambda = 0$, the major axis $R_0 =5/3$, the minor axis $a = 1/2$, and the elongation $b= 7/10$. Under these settings, we have $\alpha \approx  0.45$ and $\beta \approx 0$.

\begin{figure}
    \centering
    \includegraphics[width=0.5\linewidth]{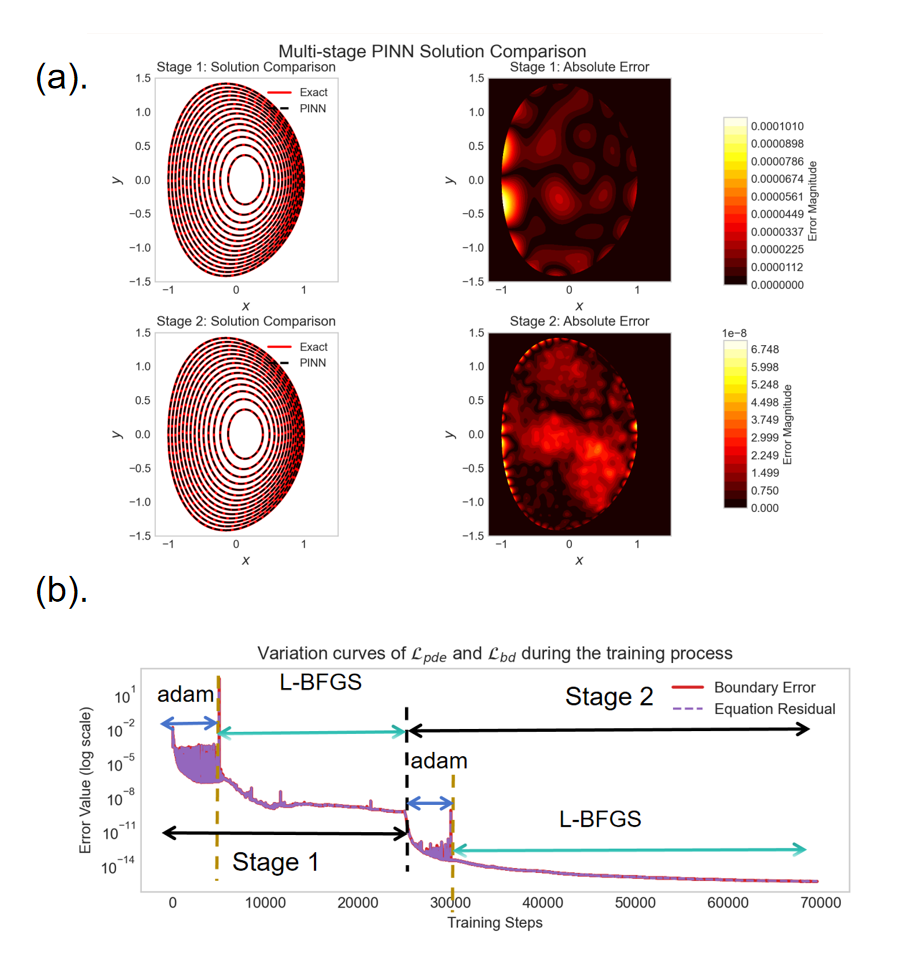}
    \caption{Solution results of the Solov\'{e}v equilibrium with triangularity parameter via Physics-Informed Neural Networks. (a) Solution results of multi-stage neural PINNs (left) and errors compared with analytical solutions (right). (b) Variation curves of $\mathcal{L}_{pde}$ and $\mathcal{L}_{bd}$ during the training process.}
    \label{fig:toroidal_device}
\end{figure}

Figure \ref{fig:toroidal_device}(a) presents the results of training with the first-stage neural network and the second-stage neural network. When comparing the solution of the GS equation obtained by the first-stage neural network with the analytical solution, the error can reach the order of $O(10^{-4})$; while the error between the result of the GS equation solved by the second-stage neural network and the analytical solution of the GS equation can reach the order of $O(10^{-8})$.

Figure \ref{fig:toroidal_device}(b) shows the variation curves of $\mathcal{L}_{\text{pde}}$ and $\mathcal{L}_{\text{bd}}$ during the training process of the neural network. Under the Adam optimizer and LBFGS optimizer, the $\mathcal{L}_{\text{pde}}$ and $\mathcal{L}_{\text{bd}}$ of the first-stage neural network can reach the order of $O(10^{-9})$, and those of the second-stage neural network can reach the order of $O(10^{-16})$. The structure and parameters of the neural network used in this example are the same as those in the first example. The test set consists of 1000×1000 uniformly distributed grid points within a rectangular region, with points outside the boundary area removed by means of masking.

\section{Conclusion and outlook}

In this study, a multi-stage PINNs is employed to solve the GS equation under various scenarios with analytical solutions. The error between the obtained results and the analytical solutions can reach the order of $O(10^{-8})$, achieving high-precision solving of the GS equation successfully. Compared with traditional numerical solution methods, PINNs can perform derivative calculations through automatic differentiation, eliminating the need for steps such as equation discretization and mesh generation, thus simplifying the calculation process. After the training of PINNs is completed, the computation time of the equation is extremely fast, which is expected to realize real-time and high-precision solving of the GS equation.

This paper has initially explored the capability of the multi-stage PINNs in solving the GS equation, but there is still room for improvement. Future work should focus on exploring the parameterized GS equation using the multi-stage PINNs and how to solve the GS equation under experimental conditions, so as to provide a powerful tool for subsequent experimental explorations.

\section{Acknowledgments}

This work was supported by the National Natural Science Foundation of China under Grant No. 12275142.

\section{Data availability}

The data that support the findings of this study are available
from the corresponding author upon reasonable request.

\bibliographystyle{elsarticle-num}
\bibliography{simple}

\end{document}